\documentclass[12pt]{iopart}

\usepackage[english]{babel}
\usepackage{graphicx}
\expandafter\let\csname equation*\endcsname\relax
\expandafter\let\csname endequation*\endcsname\relax

\usepackage{amsmath}
\usepackage{iopams}

\begin{document}

\title{The true reinforced random walk with bias}

\author{E Agliari$^{1,2,3}$, R Burioni$^{1,2}$ and G Uguzzoni$^{1,2}$}
\address{$^1$Dipartimento di Fisica, Universit\`{a} degli Studi di Parma, viale G. Usberti 7, 43100 Parma, Italy}
\address{$^2$INFN Gruppo Collegato di Parma, Italy}
\address{$^3$Physikalisches Institut, Universit\"{a}t Freiburg, Hermann-Herder-Strasse 3, 79104 Freiburg, Germany}

\ead{guido.uguzzoni@fis.unipr.it}

\begin{abstract}
We consider a self-attracting random walk in dimension $d=1$, in presence of a field of strength $s$, which biases the walker toward a target site. We focus on the dynamic case (true reinforced random walk), where memory effects are implemented at each time step, differently from the static case, where memory effects are accounted for globally. We analyze in details the asymptotic long-time behavior of the walker through the main statistical quantities (e.g. distinct sites visited, end-to-end distance) and we discuss a possible mapping between such dynamic self-attracting model and the trapping problem for a simple random walk, in analogy with the static model. Moreover, we find that, for any $s>0$, the random walk behavior switches to ballistic and that field effects always prevail on memory effects without any singularity, already in $d=1$; this is in contrast with the behavior observed in the static model.
\end{abstract}


\submitto{\NJP}

\maketitle

\section{Introduction}
The reinforced random walk is an interesting example of stochastic process with memory, defined as the statistical problem of a random walker which is attracted towards the sites it has already visited. It represents the attractive counterpart of the well-known self-avoiding walk, which is biased away from the already visited region. Memory of this type is often present in physics, in ecology, in search strategies and in biological systems:  in all these cases, the walkers modify the environment  they move on in a  local manner, by leaving signals, trails or concentrations of substances which bias their subsequent motion \cite{celani2010bacterial, pemantle2007survey, othmer1997aggregation, codling2008random, zhang2010optimality, angstmann2011continuous, agliari2005random}.
The process with memory can take place in physical space, like in bacterial motion, or on a theoretical network, as it happens in models of genetic mutations in sequence spaces \cite{deBoer1992pattern, berg1993random, wylie2011biophysical}.

The memory effect of the true reinforced random walk is \textit{locally} incorporated in the interaction between steps, and can be implemented applying several rules \cite{pemantle2007survey}. For instance, one can include the  attractive part in the dynamics by increasing the weight for the jump probability towards a given site, according to the number of times the site has been visited \cite{davis1990reinforced, GrassbergerNJP2009}. In this case, the resulting self-attracting walk collapses in any finite dimension and, at long times, the walker is confined to move on a finite set of sites. A more interesting way to account for memory effects is by letting the jump probability to a given site be $\sim \exp(k u)$, where $k=1$ for sites already visited at least once and $k=0$ for the others. This model, called the one-step (true) reinforced random walk,  is related to the Donsker-Varadhan  Wiener sausage problem \cite{donsker1975asymptotics, donsker1979number, mehra2002transition} and its properties have been studied in details in \cite{GrassbergerNJP2009, sapozhnikov1994self, ordemann2000swelling, ordemann2001structural, jimenez2003epidemic}. In particular, it was shown that in high dimensional lattices ($d \geq 2$) it exhibits a non-trivial phase transition between a diffusive and a collapsed phase as a function of the parameter $u$. This is evidenced by, e.g. the behavior of the mean-squared displacement $x_t$ and of the mean number $S_t$ of distinct sites visited up to time $t$, which change in correspondence to a critical value $u_c$. Indeed, $S_t \sim t $  for $u < u_c$ (with a logarithmic correction for $d = 2$), but as $t^{d/(d + 1)}$ for $u > u_c$. In $d=1$, no phase transition has been evidenced and $S_t$ scales as in normal diffusion,  $S_t\sim x_t \sim \sqrt{t}$.

Beyond such \textit{truly dynamical} approach for modeling self-attracting random walks, a so-called \textit{static model} has been defined. The static model - sometimes confused with its dynamic counterpart - actually describes the statistics of self-attracting chains: each possible path is considered as a particular configuration within the space the process is embedded in, and the weight of each instance is \textit{globally} assigned according to the self-attracting interaction. As pointed out in the case of the self-avoiding walks \cite{Amit1983asymptotic}, static and dynamical self-attracting random walks differ considerably and their long-time and critical behavior do not coincide.

When a field biasing the walker toward a given direction and/or point is present, a few results have been obtained. For instance, the model with bias has been studied by \cite{Mehra-2002PhysD,Redner-1983PRL,Stanley-1983PRL} in the case of static memory, evidencing the presence of a second (first) order phase transition in $d=1$ ($d\geq 2$) as a function of the bias parameter, between a ballistic regime, where the displacement $x_t$ scales linearly with time, $x_t/t \rightarrow v>0$, and a  sub-ballistic regime, where  $x_t \sim t^\alpha$, with $\alpha<1$. These results have been obtained exploiting the mapping with a model of random walk diffusing on a medium displaying a uniform distribution of traps \cite{Mehra-2002PhysD, Ioffe-2010BrazJProbStat}. However, this powerful mapping does not hold when memory effects are accounted for dynamically, so the true reinforced random walk still presents many open problems.

Here we shall concentrate on the 1D version of the true one-step reinforced random walk, accounting also for the presence of a bias; the bias is taken to point towards a given site called target. Indeed, while already exhibiting a non-trivial behaviour, the 1D model is amenable to an analytical treatment for the main statistical quantities of interest in the dynamical process. Such results may be useful in deepening the considerable controversy and confusion generated by the reinforced random walk despite its simple definition. In particular, we consider  the number of distinct sites visited after a time $t$ and its growth rate, the mean displacement and the mean time spent on the border of the (instantaneous) span and we provide an analytic expression of their asymptotic behavior in very good agreement with extensive numerical simulations. As we will show, in the presence of a bias with strength $s$, the random walk behavior switches to ballistic, i.e. $x_t \sim S_t \sim t$, namely, in $d=1$, field effects prevail on memory effects for any $s>0$, without any singularity. This is in contrast with the behavior observed in the static model  \cite{Mehra-2002PhysD, Ioffe-2010BrazJProbStat}. Finally, exploiting the results obtained in the asymptotic region, we obtain a formal expression for the probability of realizing a given path, highlighting its explicit dependence on $S_t$, $x_t$ and on the number of times the random walk has sojourned on the borders of the (instantaneous) spanned region. This allows us to properly extend to the dynamic model the mapping with the trapping problem, already exploited for the static model.

The paper is organized as follows. In Sec.~2 we provide the reader with the main definitions and in Sec.~3 we describe the mapping with the trapping model. Then, we analyze in details the main statistical quantities, such as, in Sec.~4 the number of distinct sites visited, in Sec.~5 the time spent on the edges of the spanned region and in Sec.~6 the mean time to return to the target. Finally, Sec.~7 contains conclusions and Appendices include details on analytical calculations.

\section{Definitions}
Let  $\mathcal{G}=\{ V, E \}$ be a graph with vertex set $V$ and edge set $E$. Given $i,j \in V$, the stochastic process we consider is defined by the following single-step jump probability from site $i$ to $j$:
\begin{equation}
p(j|i) =\frac{ e^{u k_t(j)}}{\sum_{j \in \Gamma(i)} e^{u k_t(j)}}, \qquad j \in \Gamma(i)
\end{equation}
with $k_t(j)={1,0}$, according to whether $j$ has or has not been visited before time $t$ respectively, $\Gamma(i)$ is the set of neighborhood sites of $i$, and $u$ is a constant, whose sign distinguishes between reinforced ($u>0$) or repulsive ($u<0$) walks. We also account for the presence of a field biasing the random walk towards ($s>0$) or away ($s<0$)  from a given site $T \in V$, hereafter called target:
\begin{equation}\label{eq:singlestep}
p(j|i)=\frac{ e^{u k_t(j)} e^{- s d_j } }{\sum_{j \in \Gamma(i)} e^{u k_t(j) - s d_j}}, \qquad j \in \Gamma(i)
\end{equation}
where $d_j$ is the chemical distance between $j$ and the target, i.e. the number of links between $j$ and the position of the target, while $s$ tunes the field effect.

In the following we consider the case of a reinforced ($u>0$) walk in the presence of an attractive target ($s>0$), defined on a $1$-dimensional lattice.
When $d=1$, the walker position at time $t$ can be denoted by a scalar quantity $x_t$, and the site where the target is located is denoted by $x_T$. By handling (\ref{eq:singlestep}) we obtain the following probability to jump from $x_t$ to $x_{t+1} = x_t \pm 1$
\begin{equation} \label{eq:prescription}
p(x_t \pm 1| x_t)=\frac{ \exp[ u k(x_{t}\pm1) - s |x_T - x_t \mp 1|]}{N_t},
\end{equation}
where $N_t$ is the proper normalization factor and $k$ dropped the redundant index $t$. Without loss of generality we can choose $x_0=0$, so that $x_t$ represents the end-to-end distance of the walk at time $t$.

\begin{figure}[htb]
\centering
\includegraphics[width=0.55\textwidth]{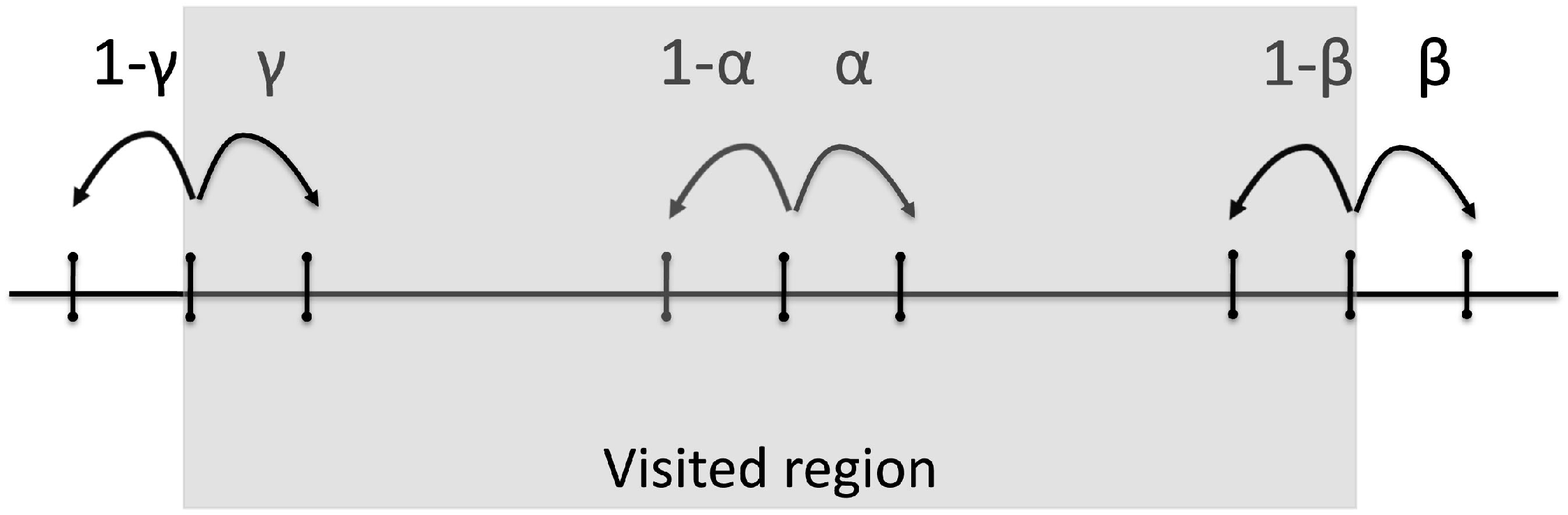}
\caption{The biased reinforced random walk in $d=1$ can be looked at as a biased random walk moving on an inhomogeneous substrate where the jump probabilities change according to the starting position of the particle: within the visited region, or on the left border, or on the right border.}\label{imm:supporto3regioni}
\end{figure}

For the analytical investigation of the process, it is convenient to focus separately on the initial transient regime (before eventually reaching the target) and the asymptotic regime (after reaching the target); these regimes correspond to a ``longitudinal field'' and to a ``central field'', respectively. In the former case, as long as $x_t + 1< x_T$, the position of the target can be ignored and the transition probability can be simplified as
\begin{eqnarray} \label{eq:prob}
&&p(x_t \pm 1 | x_t)=\frac{ \exp[ u k(x_{t}\pm1) \pm s)]}{N_t},\\
&&N_t= e^{u k(x_t + 1) + s}+e^{u k(x_t - 1) - s}.
\end{eqnarray}
More generally, we can look at the process as a biased random walk moving on an inhomogeneous substrate characterized by three regions: at the interface between the visited and the unvisited region, the single step probability assumes different expressions. Referring to figure (\ref{imm:supporto3regioni}), where the single step probabilities to move rightwards in the spanned region and at interfaces are denoted by $\alpha$, $\beta$ and $\gamma$, we have
\begin{eqnarray}\label{eq:alfabetagamma}
\alpha= \frac{e^s}{2 \cosh(s)},\\
\beta=\frac{e^{-u/2 + s}} {2 \cosh (-u/2 + s)},\\
\gamma=\frac{e^{u/2 + s}}{2 \cosh (u/2 + s)}.
\end{eqnarray}

\section{The mapping with trap models}\label{sez:StatDynMod}
According to the model described in the previous section, we assign a probability $p(j|i)$ to the walker to step from $i$ to the neighbor site $j$ during the evolution. As a result, memory effects are accounted for at each time step, that is, normalization is local.

Another common way of modeling reinforced (or repulsive) interaction between steps, is to assign to any possible realization $w$ of the walk an energy $E_{u,s}(w) = u S(w) - s x(w)$, being $S(w)$ the number of distinct sites visited and $x(w)$ the end-to-end distance displayed by the walk $w$ (here, again, $u$ and $s$ are parameters tuning the interaction). Then, the probability to realize a particular path is taken to scale exponentially with the related energy; the normalization is therefore global (see the next subsection for more details).
Models with local and global normalization conditions are also referred to as dynamic and static models, respectively \cite{ordemann2001structural}.

A comparative study of these models in one dimension \cite{Duxbury-1984JPA} has shown
that the emergent behavior depends crucially on the kind of normalization, either global or
local.
In fact, in the static model, by definition, all the paths sharing the same number of visited sites $S$ (and, in the presence of bias, sharing also the same elongation $x$) have the same probability, conversely in the dynamic model paths with the same $S$ (and $x$) can occur with different probability, depending on the number of times the walk visits the borders between different regions (see figure (\ref{imm:supporto3regioni})), as will be explained in the following.

Another point which distinguishes dynamic and static models is the possibility for the latter to be mapped into a trap model \cite{Stanley-1983PRL, Mehra-2002PhysD}; a possible extension of this mapping to the dynamic case is discussed in Sec. \ref{sec:dyn}.

\subsection{The static model}
As anticipated, in the static model \cite{ordemann2001structural,Stanley-1983PRL} an energy is associated to any path $w$; in general the energy is $E_{u,s}(w)=u S(w) -s x(w)$, where $u$ is the cost associated to a new visited site and $s$ accounts for the field \footnote{In this model the field is intrinsically longitudinal i.e., it biases the random walk towards a given direction and the strength is constant.}. Accordingly, the probability for the path $w$ of length $t$ to occur is defined as:
\begin{eqnarray}
\hat{P}_{u,s}(w)&=&\frac{e^{-E_{u,s}(w)}}{Z_{u,s}(t)}= e^{-uS(w)+s x(w)}\frac{1}{Z_{u,s}(t)},\nonumber
\end{eqnarray}
with $Z_{u,s}(t)$ the partition function, given by
\begin{equation} \label{eq:partition}
Z_{u,s}(t)  \equiv \sum_w e^{-u S(w)+s x(w)} =\sum_{S,x} \hat{W}(S,x,t) e^{-u S(w)+s x(w)},
\end{equation}
where $\hat{W}(S,x,t)$ is the number of paths of length $t$ with an elongation $x$ and $S$ distinct sites visited.\\
Starting from the partition function, we can compute the mean number of distinct sites visited $\langle S \rangle_t$, the mean displacement $\langle x\rangle_t$ and the mean square displacement $\langle x^2\rangle_t$ as
\begin{equation}
\langle S \rangle_t=-\frac{\partial}{\partial u} \ln Z_{u,s}(t),\quad
\langle x \rangle_t=\frac{\partial}{\partial s} \ln Z_{u,s}(t),\quad
\langle x^2 \rangle_t=\langle x \rangle^2_t+\frac{\partial^2}{\partial^2 s} \ln Z_{u,s}(t),\quad
\end{equation}

Now, we recall that the survival probability for a $1$-dimensional biased walk in the presence of a quenched concentration $c$ of static traps after $t$ step can be written as \cite{Mehra-2002PhysD}

\begin{equation}\label{eq:surivival}
P_{\textrm{surv}}(t) = \left(2 \cosh s  \right)^{-t}  \sum_{w} (1-c)^{S(w)} e^{s x(w)},
\end{equation}
so that, using (\ref{eq:partition}) and posing $1-c=e^{-u}$, we obtain
\begin{equation}
P_{\textrm{surv}}(t) = (2\cosh s )^{-t}   Z_{u,s}(t).
\end{equation}
Of course, for the unbiased case, starting from (\ref{eq:surivival}), we recover the well-known Rosenstock approximation
\begin{equation}
P_{\textrm{surv}}(t) = 2^{-t} \sum_{w} (1-c)^{S(w)} = \langle (1-c)^{S}  \rangle_t \approx (1-c)^{\langle S \rangle_t},
\end{equation}
where the last relation, holding in the early-time regime and for $c \ll 1$, i.e. $u \ll 1$, allows the simple estimate $Z_{u,s}(t) \approx e^{-u \langle S \rangle_t}$.
In general, through this mapping, the partition function describing the static model on the substrate $\mathcal{G}$ at time $t$ can be directly associated to the survival probability of a standard random walk on $\mathcal{G}$ at time $t$, in presence of a concentration $c= 1-e^{-u}$ of static traps.

\subsection{The Dynamical model}\label{sec:dyn}
Let us consider a particular path of $t$ steps $w=\{x_0,...,x_t\}$; exploiting (\ref{eq:prob}) we can write the probability $P_{u,s}(w)$ of realizing the path $w$ in terms of macroscopic quantities, such as the number of distinct sites visited at time $t$ and the end-to-end distance.
In fact, we have
\begin{eqnarray}
P_{u,s}(w)&=&\prod_{i=1}^t p(x_i | x_{i-1} )= \prod_{i=1}^t  \exp \big[u k(x_i) + s~ \textrm{ sign}(x_i-x_{i-1})\big]/N_i\nonumber\\
&=& e^{ -u S_t + s x_t } \left(e^{-ut} \prod_{i=0}^t N_i\right)^{-1}, \nonumber
\end{eqnarray}
where we used $\sum_{i=1}^t k(x_i)=t-S_t$.
As to the denominator, posing $k(x_i+1)=k_+$ and  $k(x_i-1)=k_-$, we obtain
\begin{eqnarray}
\prod_{i=0}^t N_i&=&\prod_{i=0}^t \exp[u (k_+ + k_-)/2]    2 \cosh [u (k_+ - k_-)/2+ s].
\end{eqnarray}
This product can be decomposed in three parts, according to whether $x_i$ corresponds to the right border ($i \in \partial_R$), to the left border ($i \in \partial_L$) or to a site within the span of the random walk ($i \in I$), namely $\prod_{i=0}^t =\prod_{i \in \partial_R} \cdot \prod_{i \in \partial_L}\cdot \prod_{i \in I}$.
Considering the values that $k_+$ and $k_-$ assume in the different regions, we obtain
\begin{eqnarray}
\nonumber
\prod_{i=0}^t N_i =  \prod_{i \in I} e^u 2 \cosh s  \prod_{i \in \partial_R}  e^{u/2} 2 \cosh(s-u/2) \prod_{i \in \partial_L} e^{u/2} 2 \cosh(s+u/2).
\end{eqnarray}
We call $t_I $, $t_R$, $ t_L$ the number of times that  the random walker has sojourned in any of the three regions, respectively. Then,  recalling that $t=t_I +t_R + t_L$, the denominator gets
\begin{eqnarray}
e^{-ut} \prod_{i=0}^t N_i&=& ( 2 \cosh s)^t \bigg[ \frac{1+ e^{2s-u}}{1+e^{2s}} \bigg]^{t_R} \bigg[ \frac{1+ e^{-u-2s}}{1+e^{-2s}} \bigg] ^{t_L}\nonumber\\
&\equiv& ( 2 \cosh s)^t \quad \textrm{C}(t_R,t_L)\nonumber.
\end{eqnarray}
Therefore, given $u$ and $s$, all the paths sharing the same span, elongation and border times, whose number shall be denoted with $W(S,t_B,t)$, are realized with the same probability
\begin{equation}\label{eq:Ppath}
P_{u,s}(w)=\frac{\exp[ -u S_t + s x_t]}{(2\cosh s)^t \textrm{ C}(t_R,t_L)}.
\end{equation}
As a consequence, the probability of realizing any of these paths is $P_{u,s}(S,t_B,t) = P_{u,s}(w) W(S,t_B,t)$. We also define $W(S, t)$ as the number of paths with span $S$ and $f(t_B|S,t)$ as the relative fraction with $t_B$ visits on border sites, so that $W(S,t_B,t) = f(t_B|S,t) W(S,t)$.

For simplicity, let us now focus on the unbiased case; this allows to neglect the quantity $x_t$ and (\ref{eq:Ppath}) simplifies into
\begin{equation}\label{eq:Ppath_dyn}
P_{u,s=0}(w)=2^{-t} e^{ -u S_t} \left(\frac{2}{1+e^{-u}}\right)^{t_B},
\end{equation}
that is, the probability of a path in the dynamic model depends on the two stochastic (not-Markovian) variables $S_t$ and $t_B = t_R + t_L$.

\begin{figure}[htb]
\centering
\includegraphics[width=1\textwidth]{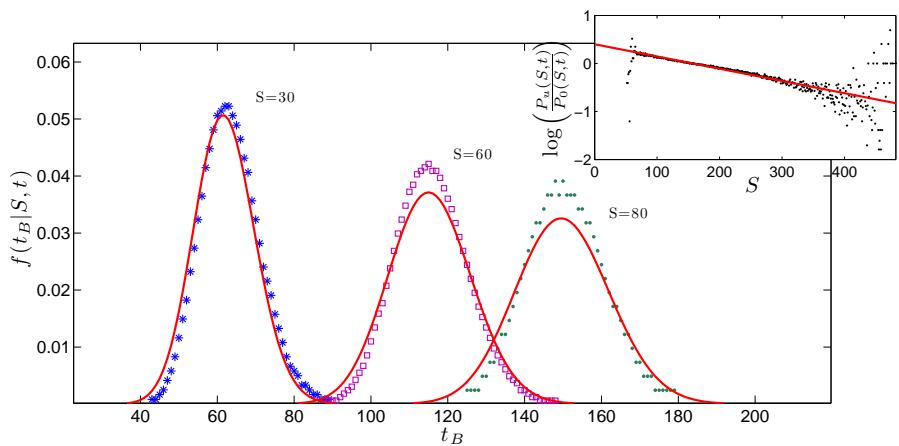}
\caption{
Main figure: fraction of paths $f(t_B| S,t)$ that after $t=1000$ steps have stayed $t_B$ times on borders, given $S=30$ ($*$), $S=60$ ($\square$), $S=80$ ($\bullet$) distinct visited sites, respectively.
Data have been fitted by Poissonian distributions. Inset: natural logarithm of $P_{u,s=0}(S,t)/P_{u=0,s=0}(S,t)$
for $t=5000$ and the linear estimate $-u'S$ (see Eq.~\ref{eq:PP}).}
\label{gr:poisson}
\end{figure}

We now calculate the probability $P_{u,s=0}(S,t)$ of realizing any arbitrary path of length $t$ which spans over $S$ distinct sites, regardless of the number of times $t_B$ it visits a border. This can be obtained by summing $P_{u,s=0}(S,t_B,t)$ over $t_B$, namely
\begin{equation}
P_{u,s=0}(S,t) = W(S,t) \sum_{t_B} P_{u,s=0}(w) f(t_B | S,t)\equiv W(S,t)  \langle P_{u,s=0}(w) \rangle_{S,t}.
\end{equation}
We note that $W(S,t) 2^{-t}$ can be looked at as the probability $P_{u=0,s=0}(S,t)$ that a simple random walk covers a span $S$ in a time $t$.
Thus, recalling (\ref{eq:Ppath_dyn}), we have
\begin{equation}\label{eq:PP}
\frac{P_{u,s=0}(S,t)}{P_{u=0,s=0}(S,t)} =  e^{ -u S} \left \langle    \left(\frac{2}{1+e^{-u}}\right)^{t_B} \right \rangle_{S,t} \approx e^{-u' S},
\end{equation}
where in the last passage we assumed $f(t_B| S,t)$ as a Poissonian with average $\langle t_B \rangle_{S,t} = 2S$, so that, exploiting the cumulant expansion, we get $u' = u - 2(1-e^{-u})/(1-e^{u})$. As shown in Fig.~\ref{gr:poisson}, this picture is in agreement with numerical results at long times (with respect to $S$), while at shorter times time-dependent corrections have to be introduced.

The expression in (\ref{eq:PP}) can be looked at as the survival probability for a random walker which has visited $S$ sites in the presence of a concentration $c=1-\exp(u')$ of traps. In the limit $u \rightarrow 0$ we get $c=0$, while for $u \rightarrow \infty$ we get $c=1$, as expected.

note that in the static case we had as variables $t$ and $S$, where the latter was canceled out via cumulants expansion.
In the dynamic case we have $t_B$ as additional variable, and it is analogously canceled out so that we are left with $P_{u,s}(S,t)$ to be compared with a survival probability of a walk where we specify both the length $t$ and the span $S$.

\section{The number of distinct visited sites}\label{sez:St}
\begin{figure}[htb]
\centering
\includegraphics[width=0.6\textwidth]{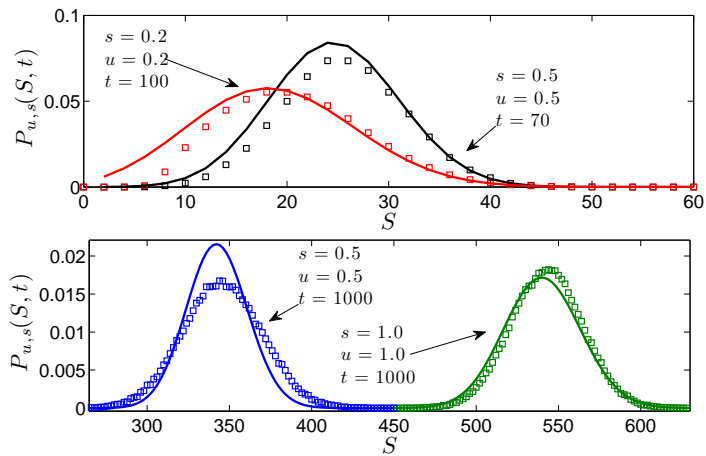}
\caption{
Upper panel: probability distribution $P_{u,s}(S,t)$ calculated at time $t=70$ and $t=100$. The agreement between simulation (symbols) and analytical estimate (from (\ref{eq:Pstdef}), solid line) is good especially for large values of $S$.
Lower panel: probability distribution $P_{u,s}(S,t)$ calculated at time $t=1000$; data from simulations (symbols) and from the estimate provided by a Poissonian distribution with parameters taken from  (\ref{eq:Taub}) and  (\ref{eq:sqmedio}).}\label{gr:t70}
\end{figure}

In this section, we estimate analytically the number of distinct sites visited, as a function of time. We recall that, restricting the analysis to the transient regime, the relative distance to the target can be neglected.
First, we focus on the distribution $P_{u,s}(S,t)$, which was introduced in the previous section and provides the probability that, at time $t$, the number of distinct sites visited is $S$. We also define
\begin{itemize}
\item $G_{u,s}(S,t)$, which is the probability that at time $t$ the span is incremented from $S-1$ to $S$, being $G(S,0)=\delta_{S,1}$,\\
\item $F_{u,s}(S,t)$, which is the probability that a walker, starting from a frontier node, is able to widen the span from $S$ to $S+1$ in exactly $t$ time steps,\\
\item $\bar{F}_{u,s}(S,t)$, which is the probability that a walker, starting from a frontier node of a span $S$, moves inwards and remains within the internal region for (at least) the following $t$ time steps.\\
\end{itemize}
Thus, the following relations hold (we drop subscripts $u,s$ to lighten the notation)
\begin{equation}\label{eq:PSt}
P(S,t)=\sum_{k=0}^t G(S,k)\bar{F}(S,t-k).
\end{equation}
As for $G(S,t)$, it satisfies the recursive equation
\begin{equation}\label{eq:GGF}
G(S+1,t)=\sum_{k=0}^t G(S,t-k)F(S,k).
\end{equation}
The two coupled equations can be treated within a generating-function formalism, being $h(\lambda) \equiv \sum_t h(t) \lambda^t$, the generating function of the arbitrary function $h(t)$. As explained in \ref{app:PS}, when the right border can be neglected (e.g., in the presence of a strong bias or at long times), the particle is likely not to return to the left border, so that in $F(S,t)$ we can drop the dependence on $S$, and we find
\begin{equation}\label{eq:Pstdef}
\tilde{P}(S,\lambda)=  \frac{1-\tilde{F}(\lambda)}{1-\lambda} \left[ \tilde{F}(\lambda) \right]^{S-1},
\end{equation}
where
\begin{equation}\label{eq:F01final}
\tilde{F}(\lambda)= \frac{2 e^{2 s} \lambda}{2 e^{2 s}+e^u-e^{2 s+u}+\sqrt{2} e^{s+u} \sqrt{1-2 \lambda^2+\cosh(2 s)}}
\end{equation}
is the generating function of the probability to extend the span on the right side.
The final formula is obtained by plugging the last expression into (\ref{eq:Pstdef}).
By anti-transforming $\tilde{P}(S,\lambda)$ we get an estimate for $P(S,t)$ which, as shown in figure (\ref{gr:t70}), is in good agreement with simulations as long as the bias is rather strong or $t$ is large. Notice that, by choosing $u=0$ and $s=0$, we get $\tilde{F}(\lambda)= \lambda/[1 + \sqrt{1-\lambda^2}]$, which consistently represents the generating function of the mean-first passage probability to a neighboring site for a simple random walk \cite{Weiss-1994Book}.

Starting from equation (\ref{eq:Pstdef}), we can also obtain the generating function of the moments of the number of distinct sites visited, that is
\begin{eqnarray}
\langle S^k\rangle_{\lambda}&=&\sum_S S^k \tilde{P}(S,\lambda)=\sum_t \lambda^t \langle S^k\rangle_t,\nonumber\\
\langle S^k\rangle_t&=&\frac{1}{t!}\frac{\partial^t}{\partial \lambda^t}\langle S^k\rangle_{\lambda} \big|_{\lambda=0}.\label{eq:smoment}
\end{eqnarray}
In particular, for the first moment we obtain
\begin{eqnarray}
&&\langle S \rangle_{\lambda} =\sum_S S \tilde{P}(S,\lambda)=\sum_S S \tilde{F}( \lambda)^{S-1} \bigg[\frac{1-\tilde{F}( \lambda)}{1-\lambda}\bigg]=\label{eq:smedio}\\
&&= \frac{ d }{ d \tilde{F} }  \left[ \sum_S \tilde{F}(\lambda)^S  \right ]  \frac{1-\tilde{F}(\lambda)}{1-\lambda}=\frac{1}{[1-\tilde{F}(\lambda)](1-\lambda)}=\nonumber\\
&&=\frac{2 e^{2 s} \lambda}{(1-\lambda)\left[e^{2 s} \left(-2+e^u+2 \lambda\right)-e^u-\sqrt{2} e^{s+u} \sqrt{1-2 \lambda^2+\rm{cosh}(2s)} \right]}.\nonumber
\end{eqnarray}

\begin{figure}[htb]
\centering
\includegraphics[width=0.6\textwidth]{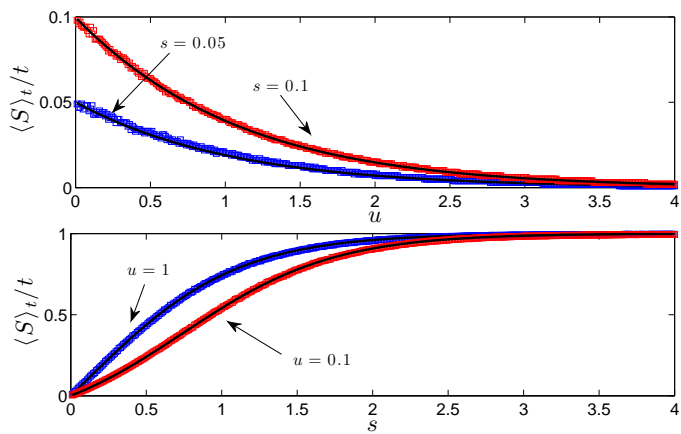}
\caption{
Growth rate for the number of distinct sites visited $\langle S \rangle_t/t$ as a function of $u$ (upper panel) and of $s$ (lower panel). Data from simulations (symbols) are compared with analytical estimates (see (\ref{eq:Taub})  and (\ref{eq:kus})).}\label{gr:slope}
\end{figure}

The previous equation can be handled out via Tauberian theorems (see e.g. \cite{Weiss-1994Book}) in order to infer the asymptotic behavior for $\langle S \rangle_t$.
In fact, $\langle S \rangle_\lambda$ can be restated as $\langle S \rangle_\lambda =(1- \lambda)^{-2} L_1[1/(1-\lambda)]$, where, posing $y=1/(1-\lambda)$:
$$L_1(y)=\frac{2e^{2s} - e^{u}(e^{2s}-1) + e^{u} \sqrt{(e^{2s} - 1)^2 + 4e^{2s}(2y-1)/y^2} }{ 2e^{2s} - y e^{u}(e^{2s}-1) + ye^{u} \sqrt{(e^{2s} - 1)^2 + 4e^{2s}(2y-1)/y^2} } $$
turns out to be a slowly varying function such that when $y \rightarrow \infty$, $L(x) \rightarrow 1/f(s,u)$, being $f(s,u)= 1 + 2 e^{u}/(e^{2s}-1)$.
Hence, we have
\begin{equation}\label{eq:Taub}
\langle S \rangle_t   \sim \frac{2t L_1(t) +t^2 L_1'(t)}{\Gamma(3)} \sim t L_1(t) = \frac{(e^{2s}-1)} {e^{2s}-1 + 2 e^{u}}~ t.
\end{equation}
This result is checked numerically in figure  (\ref{gr:slope}).

Therefore, at long times, $\langle S \rangle_t$ grows linearly with time, that is to say, the bias prevails against memory effects.
This is consistent with \cite{Ordemann-2000PRE}, where the qualitative behavior of the 1D random walk is found not to be affected by memory effects.
More precisely, the rate of growth for the number of distinct sites visited is just given by $1/f(s,u)$: it depends exponentially on $u$, without exhibiting any singularity. Indeed, for any $s>0$ the velocity turns out to be positive.
This estimate can  be compared with the result found in \cite{Mehra-2002PhysD} for the $1$-dimension static model where they evidenced a phase transition, between a phase with a zero drift velocity $v=0$ for $s<u$ and a ballistic phase with $v=\tanh (s-u)$ for $s>u$. The transition is argued to be of the second order, in the sense that $v \rightarrow 0$ continuously  and with a discontinuous second derivative.

As for the second moment $\langle S^2 \rangle_{\lambda}$, we have
\begin{eqnarray}
\langle S^2 \rangle_{\lambda}&=&\sum_S S^2 \tilde{P}(S,\lambda)=\sum_S S^2 \tilde{F}( \lambda)^{S-1} \bigg(\frac{1-\tilde{F}( \lambda)}{1-\lambda}\bigg)=\nonumber\\
&=& \frac{ d^2 }{ d \tilde{F}^2}  \left[ \sum_S \tilde{F}( \lambda)^{S}\right ]    \frac{\tilde{F}( \lambda) [1-\tilde{F}( \lambda)]}{(1-\lambda)}+\langle S \rangle_{\lambda}=\nonumber\\
&=&\frac{2 \tilde{F}( \lambda)}{ [1-\tilde{F}( \lambda)]^2 (1-\lambda)} +\langle S \rangle_{\lambda}.\label{eq:sqmedio}
\end{eqnarray}
%\frac{ \tilde{F}(\lambda)(1+\tilde{F}(\lambda)}{(1-\tilde{F}(\lambda))^2 (1-\lambda)}
 Again, we use Tauberian theorems to infer the asymptotic behaviour and we get $\langle S^2 \rangle_\lambda -\langle S \rangle_{\lambda} = (1- \lambda)^{-3} L_2[1/(1-\lambda)]$, with $L_2(y)$ converging to $2 L_1(y)^2$, when $y \rightarrow \infty$. We therefore get the asymptotic behaviour, $\langle S^2 \rangle_t \sim t^2 /[f(s,u)]^2$, which also suggests that in the limit of long $t$, the distribution for $S$ is Poissonian.

\subsection{The linear growth rate of $\langle S \rangle_t$}\label{sez:Stlinear}
\begin{figure}[htb]
\centering
\includegraphics[width=0.55\textwidth]{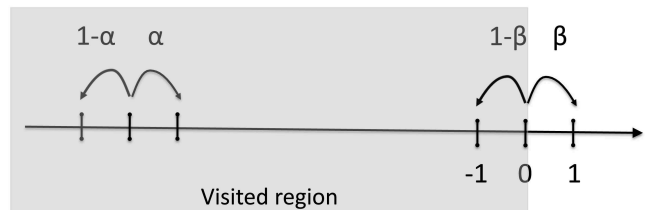}
\caption{When the random walker is far enough from the left border, we can neglect the unvisited region on the left side and just focus on how the right frontier propagates.}\label{imm:supporto_bordo_dx}
\end{figure}

A better estimate of $\langle S \rangle_t$ at small and intermediate times can be independently obtained by calculating the average time taken by the walker  to increase the extent of the visited region. In this approach we still focus only on the movement of the right edge, neglecting the movement of the other border.
More precisely, we compute the mean time taken to pass from a span of $S$ sites to $S+1$ sites. Referring to figure (\ref{imm:supporto_bordo_dx}), this corresponds to the mean first-passage time $t(0 \rightarrow 1)$. This quantity enables us to determine the rate of growth for the asymptotic law of $\langle S \rangle_t$.\\

Starting from size $0$, the random walk can either jump directly to site $1$ with probability $\beta$, or jump to site $-1$ with probability $1-\beta$; a similar position can be applied for the time $t(-1 \rightarrow 1)$ to go from site $-1$ to site $1$, thus we have
\begin{eqnarray}
&&t(0\rightarrow 1)= \beta + (1-\beta)[t(-1 \rightarrow 1)+1]\label{eq:t01right},\\
&&t(-1\rightarrow 1)= \beta \big[ t(-1 \rightarrow 0)+1\big]+(1-\beta)\big[t(-1 \rightarrow 1)+t(-1 \rightarrow 0)+1\big]\label{eq:tm11right}
\end{eqnarray}
Posing $\tau=t(-1 \rightarrow 0)$, we obtain
\begin{eqnarray}
\nonumber
t(-1 \rightarrow 1) = \frac{\tau+1}{\beta}, \;\;\;\;  t(0 \rightarrow 1) = \beta+(1-\beta)\left(\frac{\tau+1}{\beta}+1\right).
\end{eqnarray}
The explicit expression for $\tau$ is obtained in \ref{app:F01} as
$\tau= (\tanh s)^{-1},$
thus, substituting, we have
\begin{equation}\label{eq:estimate_t}
t(0 \rightarrow 1) = 1+e^u \left (\frac{1}{\tanh s}-1 \right).
\end{equation}
The mean waiting time can be related to the growth rate $v$ of $\langle S \rangle_t$, namely $d \langle S \rangle_t / d t = v \approx  t(0 \rightarrow 1)^{-1}$:
\begin{equation}\label{eq:kus}
v=\frac{1}{t(0 \rightarrow 1)} =\left[ 1+e^u \left( \frac{1}{\tanh s}-1 \right) \right]^{-1}=\frac{e^{2s}-1} {e^{2s}-1 + 2 e^{u}}.
\end{equation}
Since $t(0 \rightarrow 1)$ does not depend on time, we have $\langle S \rangle_t \approx v t$, hence recovering the same result of  (\ref{eq:Taub}), obtained through the Tauberian theorem.

We can improve the estimate of $t(0 \rightarrow 1)$ appearing in (\ref{eq:estimate_t}), taking into account also the presence of the left border, assumed as reflecting; this assumption allows to get a lower bound for $t(0 \rightarrow 1)$. In this case the time taken by the random walk to expand the visited region depends on the span length. Decomposing the first passage time in the same way as in   (\ref{eq:t01right}) and (\ref{eq:tm11right}), the only difference being the inclusion of the left border, the equations to solve are
\begin{eqnarray}
&&t(0 \rightarrow 1) = \beta + (1-\beta) \big[ t(-1 \rightarrow 1)+1 \big] , \\
&&t(-1 \rightarrow 1) =  \prod_{0|L}(-1)  \bigg\{ \beta \big[ t(-1 \rightarrow 0;L)+1\big] + (1-\beta) \big[ t(-1 \rightarrow 1) \\
&& + t(-1 \rightarrow 0;L)+1\big] \bigg\}+\prod_{L|0}(-1) \bigg\{ t(-1 \rightarrow L;0)+t(L \rightarrow -1)+t(-1 \rightarrow 1) \bigg\} ,\nonumber\\
&&t(L \rightarrow-1)=\gamma \bigg\{ \prod_{-1|L}(L+1) \big[1+t(L+1\rightarrow -1;L)\big]+\\
&&+\prod_{L|-1}(L+1) \big[ 1+ t(L \rightarrow-1)\big] \bigg\}+(1-\gamma) \bigg\{ 1+t(L-1 \rightarrow L)+t(L\rightarrow -1) \bigg\},\nonumber
\end{eqnarray}
where $L$ is the coordinate of left border set at $-S$ (see \ref{app:PS}), $\prod_{x|y}(z)$ is the splitting probability of reaching $x$ before $y$ starting from $z$ and $t(z \rightarrow x;y)$ is the conditional mean exit time, i.e. the mean first-passage time to reach $x$ from $z$ without seeing $y$ \cite{condamin2005first}. With some algebra one obtains
\begin{eqnarray}
t(0 \rightarrow 1) &= &1+ (e^u + e^{2s})(e^u + e^{-2s}) e^{-2s}e^{-2(S-1) \tanh s}+ \label{eq:new_estimate}\\
&&+ e^{u-2s} [1 + \coth (s) \; g(S,e^{\tanh s})],\nonumber
\end{eqnarray}
where $g(l,t) = [(t^{2(l-1)} -3)(t^{2l}-1) + 2l(t^2-1)] / [t (t^{2(l-1)}-1)]^2$.
We note that when $s \gg 1$, the previous equation correctly gives $t(0 \rightarrow 1) \to 1$, while for $S \gg 1$ we recover the expression in  (\ref {eq:kus}).

\begin{figure}[htb]
\centering
\includegraphics[width=0.55\textwidth]{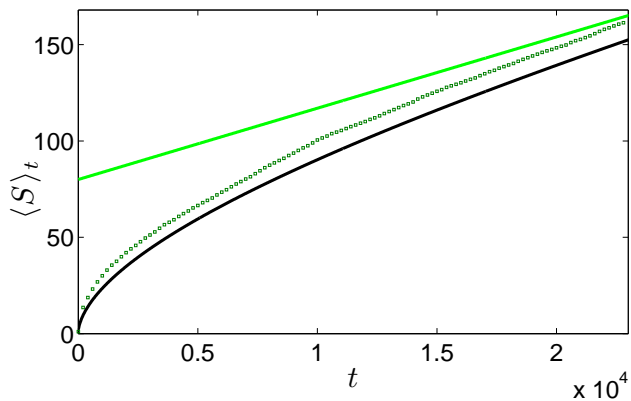}
\caption{Average number of distinct visited sites $\langle S \rangle_t$: comparison between data from simulations (symbols) and the analytical estimates where $\langle S \rangle_t \approx t / t(0\rightarrow 1)$, obtained neglecting the left border (see Eq.(\ref{eq:estimate_t}), bright line) and including the left border (see Eq.(\ref{eq:new_estimate}), dark line). Here $u=1$ and $s=0.01$.}\label{gr:rel_ns_t}
\end{figure}

This analytical estimate has been compared with data from numerical simulations in Fig. (\ref{gr:rel_ns_t}): the asymptotic behavior is nicely recovered. Moreover, the expression in Eq.~(\ref{eq:new_estimate}) with respect to the one in  Eq.~(\ref{eq:kus}),  provides a better estimate at small times, yet recovering the very same asymptotic behavior at large times.

\section{Border times} \label{sec:border}
Let us now consider the average total time $\langle t_B \rangle_t$ spent on borders and let us highlight its dependence on $\langle S \rangle_t$.
This relation was anticipated in Sec.~\ref{sec:dyn}, as it was crucial to extend the mapping between the trapping problem and the reinforced walk to the dynamic case.

Given a walk $w$, we define $t_I$ the number of times it stays on any strictly internal site and $t_{L}$ ($t_R$) the number of times that the process visits the left (right) border; of course, $t_{B} = t_L + t_R$.

Focusing on long times, from the definitions above and recalling Fig.~(\ref{imm:supporto3regioni}) and Eq.~(\ref{eq:alfabetagamma}) we can write the following relations
\begin{eqnarray}
&& t= t_{L} + t_{R}+ t_{I}, \nonumber\\
&& \langle S \rangle_t = (1- \gamma) \langle t_{L}\rangle_t + \beta \langle t_{R}\rangle_t, \nonumber\\
&& \langle x \rangle_t = (2 \gamma -1)\langle t_{L}\rangle_t + (2 \beta -1) \langle t_{R}\rangle_t + (2 \alpha -1)t. \nonumber
\end{eqnarray}
We preliminary note that a positive field ($s>0$) implies $\alpha > \beta > 1- \gamma$ (see (\ref{eq:alfabetagamma})) as well as $\langle t_R\rangle_t \gg \langle t_L\rangle_t$; therefore, we can write $\langle S \rangle_t \sim \beta \langle t_R \rangle_t$ and $\langle x\rangle_t \sim t$, from which we can infer $\langle t_R\rangle_t \sim t$, namely the propagation is ballistic, which is consistent with the above results.
On the other hand, in the absence of bias ($s=0$), we get $1-\gamma = \beta$ and $\alpha=1/2$, from which $\langle S \rangle_t \sim \langle t_B \rangle_t$ and $\langle x_t \rangle_t \sim \langle t_B \rangle_t$, consistently with \cite{Prasad-1996JPA}.

Now, we focus on the occupation of border sites, namely $t_B $, and try to highlight its connection with $\langle S\rangle_t$. We consider the unbiased case, where the random walk has probability 1/2 to move left/right within the span and probability $\phi=\exp(-u/2)/ (2 \cosh u/2 )$ to move outwards the span from a frontier site.\\

Let us denote with $\Delta S_t$ and $\Delta t_{B,t}$ the discrete differentials, given by
\begin{equation}
\Delta S_t = S_{t+1}-S_{t}, \qquad \Delta t_B = t_{B,t+1}- t_{B,t}.
\end{equation}
Since both $\Delta S_t$ and $\Delta t_{B,t}$ can assume value either $0$ or $1$, the conditional probability $P(\Delta S_t | \Delta t_{B,t-1})$ can be written in the matricial form:
\begin{equation}
P(\Delta S_t | \Delta t_{B,t-1})=
\left(
\begin{array}{cc}
P(1|1) & P(1|0)  \\
P(0|1) & P(0|0)\\
\end{array}
\right)
=
\left(
\begin{array}{cc}
\phi & 0  \\
1-\phi & 1\\
\end{array} \right) .
\end{equation}
Therefore, for the related average values we obtain
$$\langle \Delta S \rangle_t= \sum_{\Delta S_t}  \Delta S_t \sum_{\Delta t_{B,t-1}} P(\Delta t_{B,t-1}) P(\Delta S_t | \Delta t_{B,t-1} ) = P(\Delta t_{B,t-1}=1) \phi =
\phi \langle \Delta t_B \rangle_{t-1},$$
from which a linear relation between  $\langle S \rangle_t$  and $\langle t_B  \rangle_{t-1}$ follows:
\begin{equation}\label{eq:S_TB_phi}
\langle S \rangle_t= \sum_{k=0}^{t-1} \langle \Delta S \rangle_k=\phi   \sum_{k=0}^{t-2} \langle \Delta t_B \rangle_k+ \langle \Delta S \rangle_{t=0}=\phi   \langle t_B  \rangle_{t-1} +1 .
\end{equation}

\begin{figure}[htb]
\centering
\includegraphics[width=0.7\textwidth]{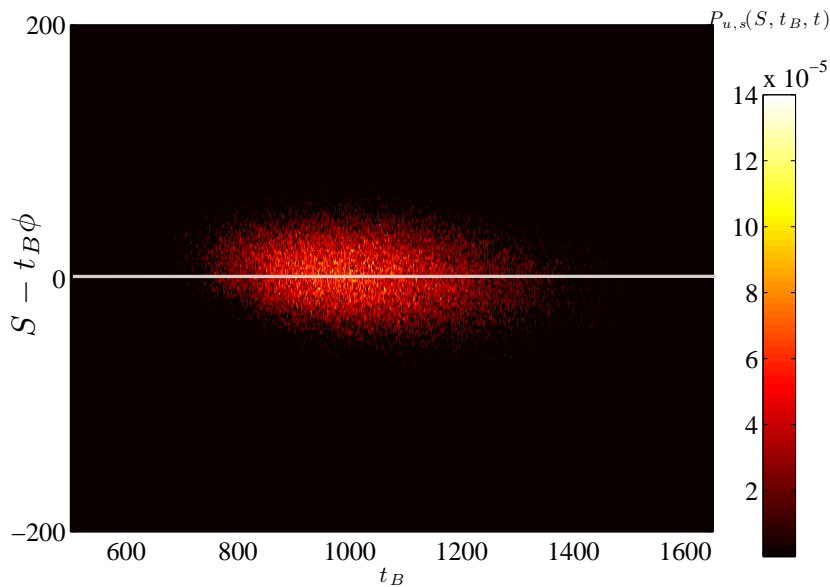}
\caption{Joint probability $P_{u,s}(S,t_B,t)$ of having the number $t_B$ of visits on borders and $S$ distinct sites visited, measured in the absence of bias $s=0$. Time is fixed at $t=100$ and $u=1$. Note that the plot is versus $t_B$ and to $S - t_B \, \phi$, in order to evidence the peak of the distribution.}\label{gr:PSe}
\end{figure}

In the bias-free case we measured numerically the joint probability $P_{u,s}(S, t_B,t)$ that the walkers has stayed $t_B$ times on any border and that $S$ distinct sites have been visited; results are represented in figure (\ref{gr:PSe}).
From this distribution we calculated $\langle S \rangle_t$, confirming (\ref{eq:S_TB_phi}).
Of curse, for $u=0$, we get $\phi=1/2$ as expected. 
 %Interestingly, for fixed $S$, $P(S, t_B)$ turns out to be a Poissonian with average $\langle t_B \rangle = \langle S %\rangle / \phi$ consistently with (\ref{eq:S_TB_phi}). Therefore, at long times, recalling that $\langle S \rangle_t %\sim \sqrt{t}$ \cite{Prasad-1996JPA}, one has $\sqrt{ \langle t_B^2 \rangle - \langle t_B \rangle^2 } / \langle t_B %\rangle \sim t^{-1/4}$; this result is also consistent with the compact exploration expected in the case of the $1$-dimensional reinforced random walks \cite{Prasad-1996JPA, ordemann2001structural, Weiss-1994Book}.

\section{Target}\label{sez:target}
In this last section we analyze the behaviour of the random walk once the target has been reached; this corresponds to the case of a random walk in the presence of a ``central field" pointing towards the point $x_T$. For simplicity we assume that the span around the point $x_T$ is symmetric and we calculate the rate of growth for the number of distinct sites visited as well as the average time taken by the random walk to return to the target.

\subsection{Mean exit time and scaling law for $\langle S \rangle_t$}
\begin{figure}[htb]
\centering
\includegraphics[width=0.6\textwidth]{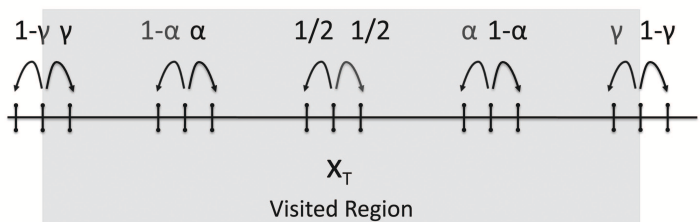}
\caption{The single step probabilities in the region around the target. In this scheme the RW is supposed to have visited a symmetric region around the target}\label{imm:supporto_TARGETtrue}
\end{figure}

After having reached the target, the single step probability follows the framework depicted in figure (\ref{imm:supporto_TARGETtrue}). In order to calculate the mean time taken by a random walk to increase the span width, we need the mean waiting time to revisit the borders. As anticipated, we take the spanned region as symmetric with respect to the target; more precisely, to simplify the notation we take $x_T=0$ and borders at $-L$ and $L$, respectively.
\begin{figure}[htb]
\centering
\includegraphics[width=0.50\textwidth]{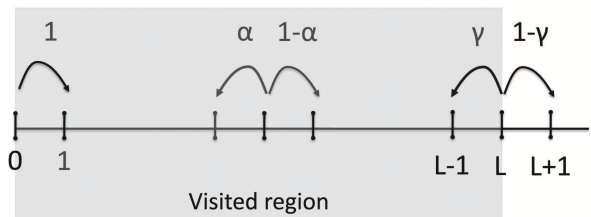}
\caption{The framework of figure (\ref{imm:supporto_TARGETtrue}) is mapped into a semi-infinite structure where, exploiting the symmetry, the target is placed at the origin and looked at as a reflective barrier.}\label{imm:supporto_TARGET}
\end{figure}
This condition simplifies the framework for the single step probability (see figure  (\ref{imm:supporto_TARGET})): we have one reflecting point in $x_T$ and a region characterized by a rightwards drift, where the probability for $x_t \rightarrow x_t+1$ is $1-\alpha$, while at the border $x=L$, the probability to move in the right direction is $1-\gamma$. Following a procedure similar to the one presented in Section (\ref{sez:Stlinear}), we can calculate the mean exit time, i.e the first passage time from $L$ to $L+1$, namely $t(L \rightarrow L+1)$. The details of the computation are shown in \ref{app:target1}, leading to:
\begin{equation}\label{eq:StDyn:tfinal}
 t(L+1 \rightarrow L+2)= 1+\frac{ 2 e^u}{1-e^{-2s}} \big[ e^{2 (s + L \tanh s )}-1 \big].
\end{equation}

\begin{figure}[htb]
\centering
\includegraphics[width=0.65\textwidth]{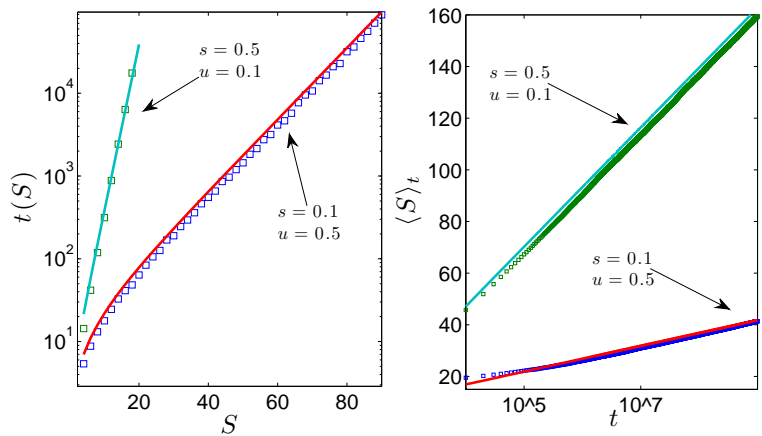}
\caption{
Left panel: mean time $t(S)$ to expand a span of width $S$ laying symmetrical around the target site. Analytical estimates (see Eq.~(\ref{eq:StDyn:tS}), solid line) are compared with data from simulations (symbols).
Right panel: Average number of distinct visited sites $\langle S \rangle_t$ as a function of time. Analytical estimates (see  Eq.~(\ref{eq:asymptS}), solid line) are compared with data from simulations (symbols). Both panels show results referring to two different choices of parameter ($u,s$), as specified}\label{gr:tS_St}
\end{figure}

From this result we can estimate the scaling law for $\langle S\rangle_t$. In fact, $t(L \rightarrow L+1)$ is the mean time the walker has to wait in order to increase the number of visited sites from $S=2L$ to $2L+1$. The time $t(S)$ corresponding to $S$ visited sites is:
\begin{eqnarray}
t(S)&\sim&\int_1^S \Delta t(S)=\int_1^S t(L \rightarrow L+1) =\nonumber\\
&=&e^{S \tanh s} \bigg( \frac{e^{3s+u-2 \tanh s}}{2 ~\sinh s \, \tanh s} \bigg)+ \frac{S}{2} \bigg( 1-\frac{e^{s+u}}{\sinh s}\bigg)+C_{u,s}, \label{eq:StDyn:tS}
\end{eqnarray}
where $C$ is a finite constant depending on $s$ and $u$. In Fig.~(\ref{gr:tS_St}) we show that the expression in  (\ref{eq:StDyn:tS}) correctly estimates the leading behavior of $\langle S \rangle_t$. In particular, in the limit $S \rightarrow \infty$, the first term in  (\ref{eq:StDyn:tS}) provides the leading contribution, therefore, at long times we can retain only this term and invert the relation to get
\begin{equation}\label{eq:asymptS}
\langle S \rangle_t = \frac{1}{\tanh s}  \log \left[\frac{t}{B(s,u)}\right],
\end{equation}
with $B(s,u)=\exp(3s+u-2 \tanh s) / (2~ \sinh s \tanh s)$.
In Fig.~(\ref{gr:tS_St}) we show the comparison with simulations. The leading behaviour of (\ref{eq:asymptS}) is in good agreement with numerical data, the difference being due to our approximations (the symmetric span,  the continue limit in (\ref{eq:StDyn:tS}) ).
Notice that the long time behaviour is dominated by the bias encoded by the factor $(\tanh s)^{-1}$, while memory effects yield small corrections encoded by the term $B(s,u)$.\\

\subsection{Mean return time to the target}
Let us now consider the mean time to return to the target, denoted as $\tau_T$. Referring to figure  (\ref{imm:supporto_TARGET}) this is given by $\tau_T =  1+ t(1\rightarrow 0)$, namely the walker takes one step to move from $x_T\equiv 0$ to $1$ and $t(1\rightarrow 0)$ to go from $1$ back to $0$.
Assuming the span $2L$, so that the distance between the target and the border is $L$, we can calculate $t(1\rightarrow 0)$ fixing the distance of the border from the target at $L$, hence imposing a reflecting barrier at $L+1$; this corresponds to a situation where $u \gg 1$ and/or $L \gg 1$. From calculations reported in \ref{app:target2}:
\begin{equation}
t(1  \rightarrow 0) = t_{\rm{exit}}(1)+ \frac{\prod_{L|0}(1)}{\prod_{0|L}(L-1)} \bigg(t_{\rm{exit}}(L-1) + \frac{2}{\gamma}+1\bigg),
\end{equation}
where $t_{\rm{exit}}$ is the exit time from an interval $L$,starting from $L-1$ and in presence of a bias towards $0$. Therefore, the mean time taken by the random walk to return to the target site is
\begin{eqnarray}
\nonumber
\tau_T&=&t(1 \rightarrow 0)+1=1+\frac{1}{\rm{tanh}(s)}\bigg[ 1-L ~ \frac{e^{\frac{1}{2}\rm{tanh}(s)}-1 }{e^{\frac{1}{2} L \rm{tanh}(s)}-1}
+ \bigg(\frac{1-e^{2\rm{tanh}(s)}}{e^{2 L \rm{tanh}(s)}-1} \bigg)\\
&&\bigg( L-1 - L ~ \frac{e^{\frac{1}{2} (L-1) \rm{tanh}(s)} -1 }{e^{\frac{1}{2} L \rm{tanh}(s)}-1}\bigg) ~\bigg]-2e^{-2 (L-1)\rm{tanh}(s)}~\big(1+e^{2s+u}\big).\nonumber
\end{eqnarray}
In the limit for $L \rightarrow \infty$,  $\tau_T $ reaches the asymptotic value
\begin{equation}\label{eq:tauT}
\tau_T= \frac{2}{1-e^{-2s}}.
\end{equation}
\begin{figure}[htb]
\centering
\includegraphics[width=0.50\textwidth]{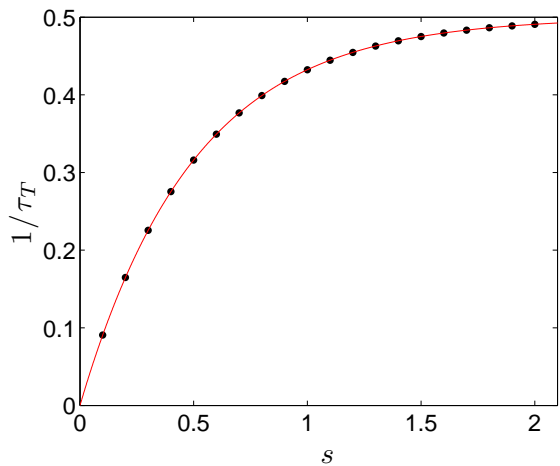}
\caption{Stationary probability $1/\tau_T$ for the biased walker to be on the target site as functions of the bias strength $s$.
The comparison between data from numerical simulations (symbols) with fixed memory interaction $u=0.1$ and the analytical estimate (solid line) given by (\ref{eq:tauT}) is very good, despite the assumptions of reflecting barrier on both borders of visited region. }\label{gr:tauT}
\end{figure}
Notice that $\tau_T$ can be related to the stationary probability for a biased random walker to be at the target site \cite{Condamin-2007PRE}. In fact, according to Kac's formula for irreducible graphs, there exists a
unique stationary probability $\pmb{\pi}$ and the mean number of steps needed to return to any point $i$, is $1/ \pi_i$.

As shown in \ref{app:target2}, a random walk embedded in a finite chain of length $L$ and moving according to the prescription (\ref{eq:prescription}) with $u=0$, admits a stationary probability with $\pi_0=(1-e^{-2s})/[2 (1 - e^{-2s(L-1)})]$. Now, assuming $L$ large, we obtain $\pi_0 \rightarrow 1 /\tau_T$. As expected, $\pi_0$ is finite when  $s>0$ and grows exponentially with $s$ up to the asymptotic value 1/2, while for $s \rightarrow 0$, $\pi_0$ is vanishing.\\

\section{Conclusions}
In this paper we studied analytically the 1D one-step reinforced random walk: at each time step the probability for the random walk to jump to a given site is $\sim \exp (ku)$, where $k=1$ for the sites visited at least once and $k=0$ for the others; a positive value for the parameter $u$, ruling memory effects, ensures that the walk is self-attractive. Since memory effects are accounted for at each time step, this model is often referred to as ``dynamic'' or ``true''.  Such models find applications in the investigation of systems where diffusing particles are able to change the environment as, e.g., in the evolution of a surface of growing aggregates \cite{krug1989microstructure}, in spatial exploration with learning \cite{pemantle2007survey} and angiogenesis \cite{codling2008random}. In our analysis we also considered the presence of a field $s$ biasing the particle towards a given direction/site.

First, we highlighted some basic differences between the dynamic and the static model. In the latter case memory effects are accounted globally: the probability of realizing a given path $w$ of length $t$ scales exponentially with the number of distinct sites visited $S$ and with its end-to-end distance $x$.
The dynamic model is by far more difficult to treat, due to the fact that memory effects are accounted locally, hence yielding a complex non-Markovian dynamics. Indeed, we found an analytic expression for the probability $P_{u,s}(w)$ to realize a particular path $w$, for a given choice of parameters $u$ and $s$, where we evidenced a dependence not only on $S$ and $x$, but also on the number of times $t_B$ that the walker has sojourned on the border sites of the (instantaneous) span.
By averaging $P_{u,s}(w)$ over all paths exhibiting the same span we found an expression which can be compared to the survival probability of a standard random walk with a given span, in the presence of a uniform concentration $c$ of traps, hence extending the mapping already proved and far exploited in the static model. Interestingly, here $c$ depends exponentially on $u$ as well as on the walker ``velocity''.

We also obtained an expression for the probability $P_{u,s}(S, t)$ of visiting $S$ distinct sites in a time $t$, in terms of generating functions and found analytically that $\langle S\rangle_t \sim \sqrt{t}$ in the absence of bias (recovering previous results \cite{Prasad-1996JPA}), and becomes ballistic with $\langle S\rangle_t \sim t$ in the presence of bias, as soon as $s>0$. Thus, memory effects induce second order corrections in this case.

Finally, we studied the joint probability $P_{u,s}(S,t_B,t)$, which for any given (large) $S$ is peaked at $t_B =  S / \phi$, where $\phi$ is the probability to move outwards starting from an edge of the spanned region.

All analytical results have been confirmed, at least in their leading terms, by numerical simulations.

\ack
This work is part of the research founded by the FIRB project RBFR08EKEV.

\appendix

\section{Distinct sites visited}
\subsection{Computation of the probability distribution P$(S,t)$}\label{app:PS}
In this section we present a procedure to calculate the generating function for the distribution $P(S,t)$ of the number of distinct sites visited $S$ up to time $t$. The latter can be formalized in terms of the probability $G(S,t)$ that a random walker increases the span width from $S-1$ to $S$ at time $t$. Indeed, the following relation holds:
\begin{equation}\nonumber
P(S,t)=\sum_{k=0}^t G(S,k)\bar{F}(S,t-k),
\end{equation}
where $\bar{F}(S,t)$ is the probability that a random walker, starting from a frontier site, does not broaden the span $S$ after $t$ steps.
Now, $G(S,t)$ satisfies the recursive equation
\begin{equation}\label{eq:GGF2}
G(S+1,t)=\sum_{k=0}^t G(S,t-k)F(S,k),
\end{equation}
where $F(S,t)$ is the probability that a random walker, starting from a frontier site, increases the span width from $S$ to $S+1$ in $t$ steps. In terms of the generating functions, equation (\ref{eq:GGF2}) becomes:
\begin{equation}
\tilde{G}(S+1,\lambda)= \tilde{G}(S,\lambda) \tilde{F}(S,\lambda).
\end{equation}
The previous finite difference equation has solution
$$\tilde{G}(S,t)=\prod_{R=1}^{S-1} \tilde{F}(R,\lambda) \tilde{G}(S=1,\lambda)=\prod_{R=1}^{S-1} \tilde{F}(R,\lambda), $$
where we have considered that $G(S=1,T)= \delta_{t,0}$, that is, $\tilde{G}(S=1,\lambda)=1$.\\

Due to the bias, the probability distribution for the position of the random walk will be peaked in the region close to the right edge and it will move away from the left edge, increasing the mutual distance. Hence, we can assume that the random walk never returns at the left edge, namely that this edge is fixed and consequently focus on the right side of the visited region. More precisely, we can fix the right border at $x=0$, so that the visited region is the half-line of the negative $x$, the single step probability towards right on this region is $\alpha=\exp(s)/2\cosh(s)$, while on the edge the probability becomes  $\beta=\exp(s-u/2)/2\textrm{cosh}(s-u/2)$ in order to account for memory effects, see figure (\ref{imm:supporto_bordo_dx}).\\

Within this framework the probability $\tilde{F}(L,k)$ is independent of $L$, and can be restated as $F(0\rightarrow1,k)$, namely the first-passage probability from $x=0$ to $x=1$ in $k$ steps.\\

\subsection{Computation of $\tilde{F}(0\rightarrow1,\lambda)$}\label{app:F01}
 For the probability $F(0\rightarrow1,t)$ the following equations hold:
\begin{eqnarray}
F(0\rightarrow1,t)&=&\beta \delta_{1,t}+(1-\beta) F(-1\rightarrow1,t-1).\\
F(-1\rightarrow1,t)&=&\sum_{k=0}^t F(-1\rightarrow0,k)F(0\rightarrow1,t-k).\nonumber
\end{eqnarray}
In the former equation the first passage probability to reach 1 from 0 is splitted in two terms according to whether the random walk moves directly to 1 or to -1, while in the latter equation the first-passage probability from -1 to 1, is decomposed in two processes: the random walk first reaches 0, then from 0 it reaches 1. The related generating functions are
\begin{eqnarray}
\tilde{F}(0\rightarrow1,\lambda)&=&\beta \lambda+(1-\beta) \tilde{F}(-1\rightarrow1,\lambda)\label{eq:F01},\\
\tilde{F}(-1\rightarrow1,\lambda)&=& \tilde{F}(-1\rightarrow0,\lambda)\tilde{F}(0\rightarrow1,\lambda)\label{eq:Fm11}.
\end{eqnarray}
By plugging equation (\ref{eq:F01}) into  (\ref{eq:Fm11}) we obtain
$$\tilde{F}(0\rightarrow1,\lambda)=\frac{\lambda \beta}{1-\lambda (1-\beta)\tilde{F}(-1\rightarrow0,\lambda)}.$$
The advantage of this formulation is that now we have to find $\tilde{F}(-1\rightarrow0,\lambda)$, which involves a simple random walk in the presence of bias; this can be found via standard techniques (see e.g. \cite{Weiss-1994Book}) as
$$\tilde{F}(-1\rightarrow 0,\lambda)= \frac{  e^{s} (\cosh(s))} {\lambda} \left(1- \sqrt{1- \lambda^2 \rm{sech}(s)^2}  \right).$$
With this result we finally get
\begin{equation}\label{eq:app:F01final}
\tilde{F}(0\rightarrow1,\lambda)= \frac{2 e^{2 s} \lambda}{2 e^{2 s}+e^u-e^{2 s+u}+\sqrt{2} e^{s+u} \sqrt{1-2 \lambda^2+\cosh (2 s)}}.
\end{equation}
In the main text $\tilde{F}(0\rightarrow1,\lambda)$ is denoted as $\tilde{F}(\lambda)$ to lighten notation.
Now, we can obtain an expression for the probability distribution of the number of distinct sites visited $P(S,t)$. The probability $\tilde{G}(S,\lambda)$ is simply $\bigg[F(-1\rightarrow0,t)\bigg]^{S-1}$, so that
\begin{equation}\label{eq:app:Pst}
\tilde{P}(S,\lambda)= \tilde{\bar{F}}(0 \rightarrow 1, \lambda)~ \big[ \tilde{F}(0\rightarrow1,\lambda) \big]^{S-1}.
\end{equation}
The probability $\tilde{\bar{F}}(0\rightarrow 1,\lambda)$ can be obtained from the first-passage probability $\tilde{F}(0\rightarrow1,\lambda)$ as
$$\tilde{\bar{F}}(0\rightarrow1,\lambda)= \frac{1}{1-\lambda} \left( 1-\tilde{F}(0\rightarrow1,\lambda) \right),$$
which plugged into  (\ref{eq:app:Pst}) yields:
\begin{equation}\label{eq:app:PS_final}
\tilde{P}(S,\lambda)=  \frac{1-\tilde{F}(0\rightarrow1,\lambda)}{1-\lambda} \left[ \tilde{F}(0\rightarrow1,\lambda) \right]^{S-1}.
\end{equation}
The final formula is obtained using (\ref{eq:app:F01final}) in (\ref{eq:app:PS_final}).\\

\section{Target}
\subsection{Computation of the mean passage time from $L$ to $L+1$}\label{app:target1}

Following a procedure similar to the one presented in \ref{app:F01}, we calculate the mean exit time, i.e the first passage time from $L$ to $L+1$, denoted as  $t(L \rightarrow L+1)$. Referring to figure ($7$), we can write
\begin{eqnarray}
&&t(L  \rightarrow L+1)= (1- \gamma)+ \gamma~ \big( t(L-1  \rightarrow  L+1)+1\big),\label{eq:app:tLLp1}\\
&&t(L-1  \rightarrow L+1)= \prod_{L|0}(L-1) ~\big[ (1-\gamma)\left( t(L-1  \rightarrow L;0)+1\right)+\label{eq:app:tm1L}\\
&&+ \gamma(t(L-1  \rightarrow L;0+1)+t(L-1  \rightarrow L+1) \big]+\nonumber\\
&&+ \prod_{0|L}(L-1)~\big[ t(L-1  \rightarrow 0;L)+t(0  \rightarrow L-1)+t(L-1  \rightarrow L+1) \big],\nonumber
\end{eqnarray}
where $\prod_{x|y}(z)$ means the splitting probability of reaching $x$ before $y$ starting from $z$ and $t(z \rightarrow x;y)$ is the conditional mean exit time i.e. the mean first-passage time to reach $x$ from $z$ without seeing $y$ \cite{condamin2005first}. After some algebra  (\ref{eq:app:tm1L}) becomes
\begin{eqnarray}
\label{eq:app:t0Lm1}
&&t(L-1 \rightarrow L+1)= \bigg[1-\gamma~ \prod_{L|0}(L-1)-\prod_{0|L}(L-1)\bigg]^{-1} \times \\
\nonumber
 && \bigg\{\prod_{L|0}(1)~ \big[ t(L-1 \rightarrow L;0)+1\big]+  \prod_{0|L}(1)~\big[ t(L-1 \rightarrow 0;L) + t(0 \rightarrow L-1) \big]\bigg\}.
\end{eqnarray}
Assuming $x=0$ as a reflection point, we can write
\begin{eqnarray}\label{eq:app:t2Lm1}
t(0  \rightarrow L-1)&=&1+t(1  \rightarrow L-1),\\
t(1 \rightarrow L-1)&=& \prod_{L-1|0}(1)~\big[t(1  \rightarrow L-1;0)\big]+\label{eq:app:t1Lm1}\\
&+&\prod_{0|L-1}(1)~\big[ t(1  \rightarrow 0;L-1)+t(0  \rightarrow L-1) \big].\nonumber
\end{eqnarray}
Substituting  (\ref{eq:app:t1Lm1}) in (\ref{eq:app:t2Lm1}) we obtain
$$
t(0 \rightarrow L-1)= \frac{1+\prod_{L-1|0}(1)~ t(1 \rightarrow L-1;0)+ \prod_{0|L-1}(1)~ t(1 \rightarrow 0;L-1)}{1-\prod_{0|L-1}(1)}.
$$
Using the last expression in  (\ref{eq:app:t0Lm1}) we have
 \begin{eqnarray}
&& t(L-1 \rightarrow L+1)= \bigg[1- \gamma~  \prod_{L|0}(L-1)- \prod_{0|L}(L-1) \bigg]^{-1}\times\\
 && \bigg\{ \prod_{L|0}(L-1) ~\big[  t(L-1  \rightarrow L;0)+1 \big] + \prod_{0|L}(L-1)~\big[ t(L-1  \rightarrow 0;L)+t(0  \rightarrow L-1) \big] \bigg\} .\nonumber
 \end{eqnarray}
 What we need in the last expression are the conditional exit times, i.e. $t(1 \rightarrow L-1;0)$  and $t(1 \rightarrow 0;L-1)$, but these involve only a simple random walk with bias and are known from the literature \cite{Redner-2001Book}. After some algebra we obtain
$$
t(L-1 \rightarrow L+1)=\exp \bigg( u/2-2s-2(L-1) \tanh s \bigg)~ \bigg( \frac{\cosh (s+u/2)}{\sinh s}\bigg).
$$
The last expression has to be substituted in  (\ref{eq:app:tLLp1}) to obtain the mean exit time from a support of $2L$ distinct visited sites
\begin{equation}\label{eq:app:tfinal}
 t(L+1 \rightarrow L+2)= 1+\frac{ 2 e^u}{1-e^{-2s}} \bigg( e^{2 (s + L\, \tanh s)}-1 \bigg).
\end{equation}

\subsection{Computation of the mean return time to the target}\label{app:target2}

Here, we derive the mean time to return to the target: looking at figure  (\ref{imm:supporto_TARGET}), this requires one step (from site $0$ to site $1$) together with the mean number of steps to return in 0 starting from 1. Hence, we distinguish between the paths which reach the border $L$ before returning to the target and those that do not. Accordingly, we have
\begin{eqnarray}
t(1  \rightarrow 0)&=& \prod_{0|L}(1)~ t(1  \rightarrow  0;L)+ \prod_{L|0}(1) ~\bigg[t(1  \rightarrow  L;0)+t(L  \rightarrow  0)\bigg]\label{eq:apptar:t10},\\
t(L  \rightarrow 0)&=&\gamma~\bigg[ t(L-1  \rightarrow  0)+1\bigg]+(1-\gamma)~\bigg[ t(L \rightarrow  0)+2\bigg]\label{eq:apptar:tL0}.
\end{eqnarray}
The last equation accounts for a reflecting barrier at $L+1$, so that the walk reaches $L+1$ and at the next step it returns to $L$ with probability $1$.
The closure of the system is due by the following equation for $ t(L-1  \rightarrow  0)$:
$$ t(L-1  \rightarrow 0)= \prod_{0|L}(L-1)~\bigg[t(L-1  \rightarrow  0;L)+1\bigg]+ \prod_{L|0}(L-1)~\bigg[t(L -1 \rightarrow  L; 0)+t(L  \rightarrow 0)\bigg],$$
which plugged in the equation (\ref{eq:apptar:tL0}), after some algebra, gives
$$t(L  \rightarrow 0)=\frac{\gamma~ \bigg[ \prod_{0|L}(L-1)~t(L-1  \rightarrow  0;L) + \prod_{L|0}(L-1)~t(L -1 \rightarrow  L; 0)\bigg]+2-\gamma}{ \gamma  \prod_{0|L}(L-1) }.$$
Now, the expression in the square bracket is the unconditional exit time from an interval $L$ in presence of a bias towards $0$ starting from $L-1$, namely $t_{\textrm{exit}}(L-1)$. Substituting  in equation (\ref{eq:apptar:t10}) we obtain the following expression:
$$t(1  \rightarrow 0)=t_{\rm{exit}}(1)+ \prod_{L|0}(1)~t(L  \rightarrow  0)=t_{\rm{exit}}(1)+ \frac{\prod_{L|0}(1)}{\prod_{0|L}(L-1)} \bigg(t_{\rm{exit}}(L-1) + \frac{2}{\gamma}+1\bigg).$$

By using the explicit expression for the splitting probabilities found in \cite{condamin2005first}, we obtain
\begin{eqnarray}
\nonumber
&&\tau_T=t(1 \rightarrow 0)+1=1+\frac{1}{\rm{tanh}(s)}\bigg[ 1-L ~ \frac{e^{\frac{1}{2}\rm{tanh}(s)}-1 }{e^{\frac{1}{2} L \rm{tanh}(s)}-1}
+ \bigg(\frac{1-e^{2\rm{tanh}(s)}}{e^{2 L \rm{tanh}(s)}-1} \bigg)\times\\
&&\bigg( L-1 - L ~ \frac{e^{\frac{1}{2} (L-1) \rm{tanh}(s)} -1 }{e^{\frac{1}{2} L \rm{tanh}(s)}-1}\bigg) ~\bigg]-2e^{-2 (L-1)\rm{tanh}(s)}~\big(1+e^{2s+u}\big).
\end{eqnarray}
This time can be related to the stationary probability of a random walk moving on a finite chain of length $L+1$ and with probability to move leftwards (rightwards) given by $p_{-} = e^{s} / (e^s + e^{-s})$ ($p_{-} = 1- p_{+}$). The overall process is described by the tridiagonal transition matrix 
\begin{equation}
T =
 \begin{matrix}
  0 & p_{-} & 0 & \cdots & \cdots & 0 \\
  1 & 0 & p_{-} &\cdots & \cdots & 0 \\
  0 & p_{+} & 0 & \cdots & \cdots & 0 \\ 
  \vdots & \vdots  & \ddots  & \ddots & \vdots & \vdots  \\
  0 & \cdots & \cdots & p_{+} & 0 & 1 \\
  0 & \cdots & \cdots & 0 & p_{+} & 0 
 \end{matrix},
\end{equation}
where we properly accounted for boundary conditions. The stationary probability $\pmb{\pi}$ can be looked at as the eigenvector satisfying $T \pmb{\pi}=\pmb{\pi}$, with $\sum_{i=0}^L \pi_i = 1$ and $0 < \pi_i < 1, \forall i \in [0, L]$. One finds
%\begin{eqnarray}
%\pi_i &=& \tilde{\pi} \left( \frac{p_+}{p_-} \right)^{i-1}, \;\;\;\ i \in [1,L-1] \\
%\pi_0 &=& \tilde{\pi} p_-  \\
%\pi_{L} &=& \tilde{\pi} \left( \frac{p_+}{p_-} \right)^{L-2} p_+,
%\end{eqnarray}
%with the normalization factor $\tilde{\pi} =(2 p_- - 1) \{ 2 p_-^2 [1- (p_+/p_-)^{L-1} ] \}^{-1}$.
\begin{eqnarray}
\pi_i &=& \tilde{\pi} e^{-2s(i-1)}, \;\;\;\ i \in [1,L-1] \\
\pi_0 &=& \tilde{\pi} e^s/(e^s + e^{-s})  \\
\pi_{L} &=& \tilde{\pi} e^{-2s(L-2)} e^{-s}/(e^s + e^{-s}),
\end{eqnarray}
with the normalization factor $\tilde{\pi} =[1-e^{-2s}]/[ 2 e^s [1 - e^{-2s(L-1)}]]$.
In particular, 
$$
\pi_0 = \frac{1 - e^{-2s}}{2 [1 - e^{-2s(L-1)}]}.
$$

\section*{References}
\bibliographystyle{iopart-num}
\bibliography{RRW}

\end{document}